\documentclass[12pt,preprint]{aastex}

\def\etal{et al.\rm}

\begin{document}
 
\title{Microlensing of a Ring Model for Quasar Structure}

\author{ Rudolph Schild\altaffilmark{1}, Viktor Vakulik\altaffilmark{2}}

\altaffiltext{1}{Center for Astrophysics, 60 Garden Street, Cambridge, MA
02138, U.S.A.}

\altaffiltext{2}{Institute of Astronomy of Kharkov National University, 
Sumska Str. 35, Kharkov, 61022, Ukraine}


 
\begin{abstract}

Microlensing observations of the Q0957+561 A,B have consistently shown
evidence for structure in the quasar that has not been evident in available
microlensing models, where the luminous source has been consistently 
modeled as a single large round structure. We show that the microlensing
features can easily be reproduced by a luminous quasar model motivated by
observations; a luminous inner accretion disc edge and outer ring-shaped
structures where the emission lines form. Such a model can explain all of
the features known from 24 years of Q0957 microlensing observations.

\end{abstract}
\section{Introduction}\label{sec:Intr}

An early dream of the community studying microlensing of quasars was to
actually learn something about the quasar's structure (Chang \& Refsdal,
1979). At the time, the
quasar was presumed to have an accretion disc which should be luminous,
and outer structure, perhaps a flock of clouds,
where the emission lines formed, but not presumed to be
a source of optical continuum.

With the discovery of the first time delay in Q0957 (Schild \& Cholfin
1986) it was immediately evident that microlensing was observed (Grieger,
Kayser, and Refsdal, 1988) and that reasonable values of the 
quasar's luminous structure were measured, based upon these models 
consisting of
a simple filled circular source characterized by one parameter, the
diameter.

However the microlensing observations that soon followed immediately showed
a problem with this over-simple model. The microlensing review of the first
10 years of data by Schild \& Smith (1991) already found that ``fine
structure is also apparent''. This motivated Schild and colleagues to
intensively monitor the system to determine an accurate time delay and to
further define the fine structure in the microlensing.

By 1996, observations on 1000 nights were compiled by Schild (1995)
and analyzed by
statistician David J, Thomson (Thomson \& Schild, 1996; Schild \& Thomson, 
 1997), with several 
remarkable conclusions. The time delay was ill-defined even with vast
amounts of high quality data, and a rapid microlensing was usually present,
with 1\% amplitudes on a time scale of days. Structure in the autocorrelation  
plots of the two images taken individually indicated that the source quasar
evidently had structure on size scales of  10 and 200 light days
(observer's clock), a factor of 10 to
100 larger than presumed accretion disc sizes. Analysis of brightness
fluctuations from a historical record by Schild (1996) bolstered the case
for large quasar structure, which was inferred to be dominated by ``rings
or clouds with a 10\% filling factor.'' He further inferred that ``Such
structure may need to be taken into account to explain the observed
high-frequency microlensing.''

By the year 2000, a new model of quasar structure was advanced 
by Elvis (2000) to
explain the complex variety of emission and absorption line properties
observed in quasars. This new unification model features two large
ring-like structures that are actually hyperbolic in cross section. The
structures contain outflows responsible for the blue-shifted broad emission
lines long known in quasar spectra, and different types of quasars were
understood as resulting from different viewing angles of these structures.
Because the sizes of these structures are comparable to the sizes already
inferred by Schild (1996) for the continuum source, an empirical model of
quasar structure now exists that is significantly different from the
popular simple accretion disc model.

In addition to the outer structures identified by Elvis, there must also be a
luminous inner edge of the accretion disc, as discussed by Colley et al
(2003). The size of this structure, about 10 light days (observer's clock), 
has already been inferred by these
authors, and it must somehow contribute to the total luminosity and
structure. 

It has long been known (Wills, Netzer, \& Wills, 1985)
that quasar energy distributions are not well
described by a simple thermal law, as is illustrated in the composite
spectral energy distribution of Elvis et al (1994). This has frequently
been described as a broad ``big blue bump'' and a ``small bump''.
Different quasars seem to differ in the relative luminosities of these
continuum sources.
Thus the existence of discrete structures in a quasar has long been
inferred observationally.

The purpose of the present report is to take this observational model
seriously and examine how such a model will be microlensed, for comparison
with microlensing observations already published. Microlensing calculations
to date (Young 1981, Kayser, Refsdal, \& Stabell, 1987, 
Schmidt \& Wambsganss (1998 and earlier
references) have faced a vexing dilemma; for luminous sources modeled as a
simple disc, small models that can produce the rapid brightness
fluctuations observed produce large brightness fluctuations never observed
in any quasar (lensed or not known as a gravitational lens system), whereas
large models producing smaller brightness fluctuations cannot produce the
rapid effects observed.

In the pages to follow, we will show that the empirical model based upon
the Elvis (2000) model of quasar structure, the double components of 
the spectral
energy distribution, and the autocorrelation peaks found in the brightness
curves can be combined to produce microlensing brightness curves similar to
the ones observed. While it might be argued that of course a complex
structure produces more parameters that will allow fits to more complex
data, we reply that observational results strongly constrain most of the
newly introduced parameters, and our models reproduce all aspects of the 
observed microlensing without parameter fitting.

\section{Parameter Estimates for the Empirical Quasar Model} 

According to the Elvis (2000) quasar unification model, most of the 
continuum and
all the emission lines originate in large ring-like structures. A cut
through the structure passing through the black hole produces conic
sections of revolution as illustrated in Figures 3 and 4 of Elvis, 2000.
The radius of this structure is given as $10^{16}$ cm for a Seyfert galaxy
like NGC5548. In our much larger radio source quasar, we presume that the
structure scales approximately as the Schwarzschild radius of the black
hole, which is a linear function of the mass. So we would estimate a factor
20 larger size scale for our Q0957 radio source quasar.

An observational estimate of this size comes directly from the
autocorrelation analysis of the quasar brightness history. Thomson and
Schild (1996) (see also Schild 1996) found autocorrelation peaks in the
quasar's brightness record, and interpreted these as indicative of
structure on scales $2 * 10^{17}$ cm. Note that these dimensions are a factor
of 10 to 100 larger than quasar sizes ordinarily associated with quasar
accretion discs (Schmidt and Wambsganss, 1999; Refsdal et al, 2000).
For our cosmological model we adopt an Einstein-de-Sitter universe with a
Hubble Constant of 60 km/sec/Mpc, and we adopt Ds/Dd = 2 to make our
calculations applicable to an ``average'' lens system, where the actual
ratio for Q0957 is 1.41, (the model is quite insensitive to this choice). 
For our physical quasar model,
we have adopted a single ring of radius $2 * 10^{17}$cm and a
radial thickness of $2 * 10^{16}$cm. The radius estimate follows from
simplistic arguments attributing the observed lags to quasar luminous
structures that probably reprocess radiations emitted near the black hole
event horizon, but are observed at Earth with a lag corresponding to the
extra time needed to traverse the quasar structure region before being
re-emitted as UV radiation in our direction (the lag times are reported
herein are as measured in the observer's frame)
The radial thickness is determined from a
Refsdal-Stabell (1991, 1993, 1997) argument, where the amplitudes of
brightness fluctuations measured over the past 100 years can be used to
infer the total area of the light emitting surface for large sources.
While some small modifications to the theory might be required for
ring-shaped sources rather than uniformly illuminated accretion discs, we
believe that the Refsdal-Stabell theory provides a good starting point for 
our calculation. We will see that this estimate indeed reproduces the
observed fluctuations on the relevant time scale of decades.

Thus we have simplified the somewhat complex outer structure 
of the Elvis (2000)
model as a simple ring with a radius and radial thickness given above.
There is also a small correction for the ellipticity of the observed ring,
where our calculation assumes a round ring. It would be logical to assume
that the quasar is seen inclined by approximately 45 degrees, making a
circular ring appear ellipsoidal in projection.

In addition to the large ring described above, there must be a smaller
luminous ring at the inner edge of the accretion disc (facing the black
hole). Justification for this structure and an estimate of its size have
been given by Colley et al (2003). The existence of complex quasar
structure has long
been inferred from observation of ultraviolet energy distributions of
quasars; these have identified a power law continuum and a ``blue bump''
which we attribute to this small accretion disc ring. Thus we estimate the
radius of this ring in Q0957 as $10^{16}$ cm (Colley et al 2003). Our
brightness measurements have been made with an R filter having effective
wavelength 6400 Angstroms; for a redshift of z = 1.41, the proper
wavelength becomes 2655 Angstroms, where the amplitude of the blue bump is
typically 20\% of the total brightness. This is also compatible with the 
autocorrelation plot of the two quasar images, Fig. 3 of Schild (1996); here
the autocorrelation peaks defining structures at 3, 50, 80, and 
230 proper days
are all of about the same amplitude, suggesting that they all contribute
about equally to the luminosity. For the thickness of this ring, we will
simply use the Schwarzschild diameter of the black hole as a best estimate,
approximately $2 * 10^{14}$ cm. Note that for a quasar seen at 45 degrees
inclination, the relevant thickness estimate is the greater of the true
accretion disc thickness and the radial depth of the luminous structure
into the accretion disc.

Thus we have come to a model of the luminous quasar structure that is
dominated by two ring-shaped structures, the inner accretion disc edge and
the outer surface where the emission lines originate. The radii, radial
extents, and brightness ratio for these structures are summarized in
section 5.

\section{Microlensing of the Large, Outer Structure}

In our modeling to follow, we will consider the microlensing of the
quasar's inner and outer structures
independently, and then infer from the
principle of superposition that the microlensing brightness curves would be
the sum of the two components. For the large outer ring, we show in Fig. 1
the microlensing expected. For a random distribution of uniform solar mass 
stars, a
pattern of diamond-shaped caustics is presumed to cross in front of the
quasar's luminosity with a transverse pattern speed of 800 km/sec. We have
modeled the overall lens with parameters appropriate to the A quasar image; 
shear = 0.2, total optical depth 0.4, divided as stellar optical depth 0.04
and continuous optical depth (or planetary microlenses) 0.36. These
parameters may be seen to be appropriate to many gravitational lenses, in
the sense that they are about average for the approximately 70 lensed
systems now known. 

In Fig. 1 we show the general view of the microlensing
brightness pattern in relation to the ring-shaped luminous quasar source,
presumed to be in relative motion. The locus of points for the center of
the ring is the straight line, and the curve above shows the
computed brightness (magnification) for the location of the ring center
below. A time scale shown at the bottom is appropriate for our adopted
transverse velocity of 800 km/sec. The amplification scale on the left
shows that relative to the unlensed quasar brightness, the observed image A
brightness is ordinarily about a factor of 3 greater. Our calculations show
amplifications varying from 2.8 to 3.7, with an rms deviation from the mean
value of 0.1,
sampled on a time scale of 20 years. Several M-shaped events are seen in
Fig. 1 at times 350, 420, and 830 years. 

These results agree well with the observations of Q0957 microlensing
reported by Schild (1996). Brightness fluctuations sampled on time scales
of decades were found to be  [0.44, 0.26] magnitudes for 
images [A,B] respectively. These estimates contain intrinsic quasar
fluctuations as well as microlensing fluctuations, and the true
microlensing is presumed to be about the 0.1 mag rms value found in the
model. It is not an accident that the modeled fluctuations match
approximately the observations; the radial thickness (area) of the 
outer ring was computed to give the amplitude of
fluctuations reported in Table 1 of Schild (1996) using the Refsdal-Stabell
(1991, 1993, 1997) theory.

The M-shaped events on 25-year time scales match almost exactly the 
event detected 
in the era 1980-1990 (Schild \& Smith 1991; Refsdal et al 2000), and even
includes the small ``dip'' that may be seen at JD 2450000 in 
Figures 9 and 16 of Pelt et al (1998). In this interpretation, the Q0957
system has passed about half way through such an event, and in the coming
decade should return to the upper plateau level before again dropping.
These M-shaped events appear to be a characteristic feature of
microlensing by a ring-shaped luminous structure at these low optical
depths and were not anticipated in the analysis of the observed Q0957
microlensing by Refsdal \etal (2000). The observed event had an amplitude
of 0.2 mag, whereas events from our present model are closer to 0.1
magnitudes and have an elapsed time of 15 years from onset to the central
``dip''. Thus the observed microlens is indicated to be slightly below
solar mass for our stated transverse velocity. The amplitude is in
reasonable agreement for the small statistical sample available,
especially since other events with the observed 0.2 mag
rise are also found in our simulation.

It would be tempting to predict that the microlensing will now continue
with a symmetrical profile such that the coming 15 years will be the
reverse of the period 1981 - 1997. However this is statistically
disfavored, because the calculation presented here is for the low optical
depth A image, and the event is probably a microlensing of the higher
optical depth B image.

In principle the outer ring could also produce fluctuations due to fine
scale structure in the ``continouos opacity'' component, which has an
optical depth of 0.36. We have modeled the continuous opacity as both a
truly continuous opacity and as a
population of planetary mass microlenses of $10^{-5}$ solar mass.
We find that the planetary mass microlensing by the outer ring would
produce fluctuations of only .001 mag, which presently would be
unmeasurable among the several larger effects predicted.

\section{Microlensing of the Inner Luminous Structure}

The microlensing of the inner accretion disc edge is expected to produce
larger amplitude and faster microlensing effects, but their amplitudes will
be diluted by the nearly constant outer ring emission, which we have shown
to be nearly constant on year- and sub-year time scales.
We show in Fig. 2
the microlensing effects expected when the luminous inner ring crosses the
diamond caustic pattern of a solar mass microlens, with the continuous opacity
again modeled as planetary mass microlenses randomly distributed.
Here we find an
excitingly new pattern that is relevant to the published brightness curves.
Numerous sharply peaked brightness profiles are predicted that have amplitudes
of approximately 30\% and durations of approximately half a year.
Numerous sharp peaks of this description have been found in the 
Q0957 brightness
record, illustrated as Fig. 2 in Schild (1996). The typical amplitudes of
the events from observations, 0.05 magnitudes, are about appropriate for cusp
profiles with peaks of 30\% but diluted by the nearly constant outer ring
luminosity which is 80\% of the total. Such brightness peaks are presumably
responsible for the broad 3.5 cycle/year maximum in the 
Fourier Power Spectrum of the
microlensing, Fig. 4 of Schild, (1996). A detailed profile of one of these
peaked brightness features is illustrated as Fig. 5 of Schild (1996).

Also seen in the Schild (1996, Fig. 5) profile is a continuous pattern of
fine structure having an amplitude of only 0.01 mag and a time scale of
only 10 days. Our model reproduces these fine features if the accretion
disc thickness parameter is taken to be $2*10^{14}$cm, which is slightly
smaller than we originally estimated from the black hole Schwarzschild
diameter. We are not concerned about issues of the stability of such fine
structure, because this illuminated region is presumably part of a larger
accretion disc structure that has already been extensively described in the
literature and is presumed stable.

We illustrate in Figure 3 a synthetisized profile from our Fig. 2
brightness curve that shows how the profiles can be easily matched to the
observations. Although the exact locations of the fine structure do not
match because they are generated by random microlensing mass distributions,
the amplitudes and widths of the features match quite well. The widths of
the features were estimated from observations to be typically 10 days, and
our calculation without parameter fitting gives widths nearly 30
days. We presume that adjustment of the transverse velocity and thickness
of the inner ring will eventually produce closer agreement.

Figure 3 also shows that the fine structure originates in the planetary
mass microlenses. In Fig. 3 the thin curve shows the microlensing pattern
for an optical depth of 0.36 as planetary mass microlenses, and the heavy
curve is for an equivalent optical depth of smoothly distributed dark
matter. We see from this figure that the fine structure
pattern (thin curve) disappears when the planetary mass microlenses are
removed. 

To emphasize the similarity between the the observed and modeled profile,
we show them side-by-side in Fig. 4. Here we see that the general shape and
structure are very similar. The exact locations of the fine structures do
not align in the modeled profile because they result from randomly
distributed planetary mass microlenses. The duration time of the observed
profile is a factor 5 shorter than in the model profile, suggesting that
the width of the inner ring has been modelled too large, but this thickness
is one of the least constrained parameters in the model. For a
statistically preferred quasar inclination of approximately 45 degrees, the
thickness parameter for the side of the inner ring closest to the Earth
is probably different than for the far side; the latter is surely lensed
by the black hole.

It is also worth noting in Figs. 3 and 4
that the fine structure pattern seems to
have about equal numbers and amplitudes of positive and negative peaks,
with a small bias toward more positive events. Our small optical depth of
0.36 may be insufficient to produce exactly equal positive and negative 
microlensing structure, which is
a well known feature of microlensing when the optical depth is near
unity (Schneider, Ehlers, and Falco, 1992, p. 343)
and which is very difficult to contrive if structure is caused by bright
points (Gould \& Escude-Miralde, 1997) or dark clouds (Schechter et al, 2002; 
Wyithe \& Loeb, 2002) in the quasar accretion disc.

To understand what is happening, it is useful to again examine Fig. 2. When
the inner luminous ring passes behind the magnification pattern for the
planetary mass microlenses far away from a diamond caustic, 
fluctuations of only 0.01 magnitudes 
are produced, but because these are swamped by the larger and nearly constant 
luminosity of the outer ring, they would be observed at an amplitude five
to ten times smaller. But as the luminous ring passes behind the
diamond caustic, the stronger peaked profiles formed by the planetary mass
microlenses in the strong shear of the solar mass microlens produces
the stronger rapid fluctuations.
 
The existence of microlensing fluctuations of such short duration and low
amplitude is still controversial because of a fundamental conundrum in the
observational program; to recognize such fluctuations the time delay must
be known within a fraction of a day, but because of the microlensing, it is
extremely difficult to determine such an accurate time delay. This
conundrum has now been broken by the QuOC-Around-The-Clock monitoring
program (Colley et al 2002, 2003) wherein 12 observatories around the globe
continuously monitored the Q0957 system for 10 days and
produced a time delay accurate to a tenth of a day. This delay has now 
been used to show a microlensing event in published data having a statistical
certainty exceeding 99.9 \% and an amplitude of 1 \% on a time scale of
half a day (Colley and Schild, 2003). The published data
may be seen in Fig. 7 of Colley and Schild (2000), where the significance
of the event was noted but time delay uncertainty at the time precluded
statistical certainty.

The observed Colley and Schild (2003) 12-hour event implies even sharper
quasar structure, or, less plausibly, a substantially larger transverse
velocity than modeled here. Detection of the longer 30-day events modeled
here will be difficult because data of accuracy better than 1
observing campaigns of weeks to months will be required; such observations
would be of the utmost importance for the detection and study of the
baryonic dark matter.

Another important aspect of these model calculations in agreement with the
observed Schild (1996, Fig. 5) profile is the asymmetry. After the peak of the
profile is past, the observed profile
asymptotically approaches a brightness level that is significantly higher
than the level of the approach side. It may be seen that this is a simple
result of the significant quasar luminosity that has passed to the region
behind the diamond caustic, which has a substantial magnification. Note
that this asymmetry allows us to catalogue the comings and goings of such
crossings; we can predict that after a few years time there should have
been a second peak followed by a return to a lower level. Similar effects
have been predicted also for solid accretion discs passing behind diamond
shaped microlensing caustic patterns, as first demonstrated by Young (1981)
and as described and explained by Kayser, Refsdal, \& Stabell (1986).

\section{Conclusions and Discussion}

The principal result of our modeling is that a double ring 
quasar luminous structure
can produce all of the Q0957 microlensing effects recorded in observations 
to date. Most
important is the conclusion that the microlensing peaks with 5\% amplitude 
and 100 day time scale are easily predicted. The signature of a population
of planetary mass microlenses would be fluctuations is the brightness
with weekly duration and equal positive and negative amplitudes of  1\%.

A devil's advocate would note that our more complex model has 5 adjustable
parameters and can of course explain any observation. The parameters
describing our double ring model are:

1. the radius of the large ring, $2 * 10^{17}$cm

2. the radial thickness of the large ring, $2 * 10^{16}$cm

3. the radius of the small ring, $1 * 10^{16}$cm

4. the thickness of the small ring, $2 * 10^{14}$cm

5. the brightness ratio of the two rings, 4:1

We note, however, that these parameters are not arbitrarily adjustable, and
have been in fact inferred from published astronomical data and analysis.

With the adoption of the double-ring quasar structure model, our
calculations appear to successfully model 10 aspects of the Q0957 brightness
history, including:

1, 2. The history of brightness fluctuations on time scales of 20 - 100
   years. The model produces fluctuations of the (1)time scale and (2)
   amplitudes of the observed fluctuations.

3, 4. The existence of brightness cusps with (3) 5\% amplitude and (4) 
   100-day duration. 

5. The asymmetry of the above profiles.

6. The 10-day (30 day) time scale of the fine brightness fluctuations.

7. The 1\% amplitudes of the fine brightness fluctuations.

8. Both positive and negative events in the fine structure. 

9. The peaks observed in the autocorrelation estimates for the observed
   brightness fluctuations.

10. The structure in the quasar's ultraviolet energy distribution.

It will obviously be of interest to undertake further modeling of this kind
to understand the sensitivity of the model to the details of the estimated
model parameters. Thus far these parameters have been derived from
observational constraints, and many are interrelated; for example,
adjusting the brightness ratio of the two rings causes changes in the
brightness effects of both rings, but this ratio is not highly adjustable
because it must also be compatible with the observed nature of the
spectroscopically measured ultraviolet quasar continuum.

In a sense, it will be noticed that the model seems robust in that it can
produce 10 measured structure parameters, with only 5 available
parameters. Because the Q0957+561 A image analyzed here is fairly typical of
lensed quasar images, the model calculations should be approximately
applicable to many such systems.

We admit that we have been somewhat loose in claiming agreement between
models and observations, even though disagreements of a factor of two or so
may sometimes be found. We find this of little concern thus far, because
the model is clearly oversimplified in many subtler respects. 
For an inclined quasar the far side of the inner ring must
suffer relativistic effects as the radiation passes the black hole and is
lensed. Moreover, the structures we model as
round rings are in reality ellipses for any reasonable quasar inclination,
and as we describe the radial
thicknesses of the rings they must also have some thickness in the z (polar)
direction. We presume that these effects introduce errors of approximately
a factor of the square-root of two.

\begin{acknowledgements}

The authors thank Prof. Anatoly Minakov and Dr. Victoria
Tsvetkova  for helpful discussions and a sustained interest
in our work.  V.Vakulik is especialy grateful to Dr.James
Bush and Prof. Kim  Morla (Pontificia Universidad Catolica
del Peru, Lima)  for their valuable financial support, which
has made the  simulation possible. Helpful discussions with Dr. Martin
Elvis are appreciated and acknowledged.

\end{acknowledgements}

\begin{figure}[t]
\caption{Microlensing of the outer
ring-shaped luminous quasar structure by identical randomly distributed
solar-mass field
stars. For each point along the line showing motion of the
center of the
quasar ring source past the high-amplification caustics, the amplification
of the light by the field stars is shown as a broken curve. For the low
optical depth presumed for the luminous baryonic matter, 0.04, the diamond
shaped caustics produced by randomly distributed stars hardly overlap, and
many M-shaped brightness profiles with 50-year durations are predicted.}

\label{fig1}
\end{figure}

\begin{figure}[t]
\caption{Microlensing of the inner ring-shaped luminous quasar structure.
As the ring and amplification pattern cross each other, for each
center of the ring on the straight line, the amplification is shown above as a
broken curve. This curve has more structural detail because of the smaller
diameter and thickness of the inner ring. The magnification pattern is
calculated for a $0.1 M{_\odot}$ star and for a random distribution of 
$10^{-5}M{_\odot}$ missing mass particles constituting 90 percent 
of the baryonic mass.}
\label{fig2}
\end{figure}

\begin{figure}[t]
\caption {Details of the expected brightness fluctuations as the inner
ring passes the magnification pattern produced by stars and
planetary mass dark matter objects. In the main plot the
amplification relative to no diamond-shaped caustic is plotted as
a function of time in years. The heavy curve shows the
amplification if the baryonic dark matter is distributed in a
smooth sheet, and the light curve shows the effects of the
equivalent amount in planetary mass objects. In the
inset, a magnified view of the peak at 7 years is shown with the
time scale expanded and the planetary mass case vertically
shifted slightly. The expected brightness is shown in magnitudes
to permit easy comparison with the observational result, Fig. 5
of Schild, 1996, for our adopted 1:4 ratio of inner ring to outer
ring brightness.}
\label{fig3}
\end{figure}

\begin{figure}[t]
\caption {Comparison of the modeled microlensing profile and published
observations. In the left panel, we repeat the curve plotted as an inset to
Fig. 3 but here with magnitude and time scales shown to permit immediate
comparison with observations. Thus dates and magnitudes are shown for 
arbitrary zero points. The heavy solid line shows the computed profile for
continuously distributed dark matter, and the lighter line shows the
effects of planetary mass dark matter objects.
The right panel is a repeat of data published as
Fig. 5 of Schild (1996) with observation dates and observed magnitudes
shown. The shaded interval is a 1 sigma error interval for the 90 
nightly average
data points on which the profile is based. Interference between the
printer's lithographic dot pattern and the original dot pattern corrupted
this error zone shading in the original publication. 
The heavy solid line is a
simple Lorentzian profile fitted to the data, and the dashed line shows a
cubic fit to the residuals.}
\label{fig4}
\end{figure}

\end{document}